\documentclass[12pt]{article}
\input epsf
\begin{document}
\thispagestyle{empty}
\begin{center}

{\Large\bf Recent tests for the statistical parton\\
\vskip 0.3cm
 distributions }
\vskip1.4cm
Claude Bourrely $^a$,  Franco Buccella $^b$ and Jacques Soffer  $^a$
\vskip0.3cm
$^a$ Centre de Physique Th\'eorique\footnote{Unit\'e propre de  Recherche 
7061}, CNRS-Luminy, \\Case 907, F-13288 Marseille Cedex 9 - France
\vskip 0.2cm
$^b$ Dipartimento di Scienze Fisiche, Universit\`a di Napoli,\\
Via Cintia, I-80126, Napoli
and INFN, Sezione di Napoli, Italy
\vskip 2cm
{\bf Abstract}\end{center}
We compare some recent experimental results obtained at DESY, SLAC and Jefferson 
Lab., with the predictions of the statistical model, we have previously
proposed. The result of this comparison is very satisfactory.
\vskip 6cm
\noindent CPT-2002/P.4451

\noindent UNIV. NAPLES DSF-24/2002
\newpage
\setcounter{page}{1}
Deep inelastic phenomena have played a crucial role in the discovery of QCD, as
the theory of strong interactions, to establish its property of being asymptotically
free \cite{pol73} and to provide the logarithmic scaling violations found experimentally.
However, concerning our present knowledge of the parton distributions at a given $Q^2$, 
the situation is far from being settled. Indeed, the very precise measurements in 
unpolarized deep inelastic scattering (DIS) for electron-nucleon 
and charged current neutrino induced reactions, yield the structure functions
$F_2^{p,n}(x,Q^2)$ and $xF_3^{\nu N}(x,Q^2)$, which involve the combinations
of parton distributions $q_i(x,Q^2) \pm  \bar {q}_i(x,Q^2)$, summed over the flavors
$i=u,d,s,...$. Therefore, due to data statistical limitations, there is some ambiguity in deriving
from these structure functions, the quark and antiquark distributions for each flavor.
In fact a flavor symmetric sea, namely $\bar d(x) = \bar u(x)$, assumed in the Gottfried sum rule 
\cite{got67}, was disproved by the NMC Collaboration \cite{nmc94} 
and one gets instead, at $Q^2 = 4 \mbox GeV^2$,
\begin{equation}
\bar d - \bar u = \int_0^1\,(\bar d(x) - \bar u(x)) dx = 0.153 \pm 0.015\, .
\label{eq1}
\end{equation}
>From a global QCD analysis of DIS, we recall that there is also some evidence 
for an asymmetry between $s(x)$ and $\bar s(x)$ in the nucleon sea \cite{FZ}.

Polarized DIS yield the measurements of the spin-dependent structure
functions $g_1^{p,n,d}(x,Q^2)$, which are combinations of
$\Delta q_i(x,Q^2) + \Delta \bar q_i(x,Q^2)$ and don't
allow to disentangle the $\Delta q_i$ and $\Delta \bar q_i$ contributions.
The original Ellis-Jaffe sum rule \cite{ell74}, which was obtained with
the assumption that only $u$ and $d$ quarks contribute to it, reads for the
proton case 
\begin{equation}
\Gamma_1^p = \int_0^1\,g_1^p(x) dx = \frac{F}{2} -\frac{D}{18}
= 0.185 \, .
\label{eq2}
\end{equation}
The EMC Collaboration discovered, nearly 15 years ago, that this sum rule 
has also a substantial defect, since they found, at $<Q^2>=10.7 \mbox {GeV}^2$, 
$\Gamma_1^p= 0.126 \pm 0.010(stat) \pm 0.015(syst)$ \cite{ash88}.
According to an interpretation of this result which was proposed earlier, the
strange quarks had a large negative contribution \cite{GR}. However the EMC result has been confirmed 
to a higher level of accuracy, for example, at $Q^2=3 \mbox {GeV}^2$, 
one finds  $\Gamma_1^p=0.132 \pm 0.003(stat) \pm 0.009(syst)$ \cite{KA}. 

Over the last few years several empirical parametrizations have been proposed with
$\Delta p(x) = P_p(x)p(x)$, without a physical interpretation of the
parameters which appear in the unpolarized part $p(x)$ and in the polynomial $P_p(x)$.
By observing that $u^+(x)$ is the parton dominating at high $x$, a
first attempt to relate unpolarized to polarized quark distributions has
been given in \cite{buc93}, suggesting for $x \geq 0.2$, where the valence 
quarks dominate,
\begin{equation}
\Delta u(x) = u(x) - d(x) \, .
\label{eq3}
\end{equation}
It leads to the following relation between the unpolarized and polarized
structure functions 
\begin{equation}
x g_1^p(x) = \frac{2}{3} [F_2^p(x) - F_2^n(x)] \, ,
\label{eq4}
\end{equation}
if one neglects the contribution of $\Delta d(x)$, which is expected to be
smaller than $\Delta u(x)$ and whose contribution is reduced by the
factor $e_d^2/e_u^2 = 1/4$. In fact, by using the available data \cite{nmc94} for the
r.h.s. of Eq. (\ref{eq4}), one predicts $\Gamma_1^p= 0.156$ at $Q^2 = 4 \mbox{GeV}^2$, which is in
reasonable agreement with the data.

The existence of the correlation, broader shape higher first moment, suggested
by the Pauli principle, has inspired the introduction of Fermi-Dirac (Bose-Einstein)
functions for the quark (gluon) distributions \cite{bour96}. After many years
of research, we recently proposed \cite{bour02}, at the input scale
$Q_0^2 = 4 \mbox{GeV}^2$
\begin{eqnarray}
x u^{+}(x,Q^2_0) &=& {AX_{0u}^{+} x^b \over \exp[(x-X_{0u}^{+})/{\bar x}]
+1} + {\tilde A x^{\tilde b} \over \exp(x/{\bar x}) +1}\, , 
\label{eq5} \\
x \bar u^{-}(x,Q^2_0) &=& {\bar A (X_{0u}^{+})^{-1}x^{2b} \over 
\exp[(x+X_{0u}^{+})/{\bar x}]+1} + {\tilde A x^{\tilde b} \over 
\exp(x / {\bar x}) +1} \, , \label{eq6}\\
x G(x,Q^2_0) &=& {A_G x^{\tilde b +1} \over \exp(x /{\bar x}) - 1} \label{eq7}
\, ,
\end{eqnarray}
and similar expressions for the other light quarks ($u^{-},
d^{+}~\mbox{and}~ d^{-}$) and their antiparticles.
We assumed $\Delta G(x,Q^2_0) = 0$ and the strange parton distributions 
$s(x,Q^2_0)$ and $\Delta s(x,Q^2_0)$ are simply related \cite{bour02} to
$\bar q(x,Q^2_0)$ and $\Delta \bar q(x,Q^2_0)$, for $q = u,d$.
A peculiar aspect of this approach, is that it solves the problem of
desentangling the $q$ and $\bar q$ contribution through the relationship 
\cite{bha01}
\begin{equation}
  X_{0u}^{+} + X_{0 \bar u}^{-} = 0 \, ,
\label{eq11}
\end{equation}
and the corresponding one for the other light quarks and their antiparticles.
It allows to get the $\bar q(x)$ and $\Delta \bar q(x)$ distributions
from the ones for $q(x)$ and $\Delta q(x)$.
 
By performing a next-to-leading order QCD evolution of these parton distributions, we
were able to obtain a good description of
a large set of very precise data on $F_2^p(x,Q^2), F_2^n(x,Q^2), xF_3^{\nu N}(x,Q^2)$
and $g_1^{p, d, n}(x,Q^2)$ data, in correspondance with the {\it eight} free parameters :
\begin{equation}
X_{0u}^{+} = 0.46128,~ X_{0u}^{-} = 0.29766,~X_{0d}^{-} = 0.30174,~X_{0d}^{+} = 0.22775 \, ,
\label{eq8}
\end{equation}
\begin{equation}
\bar x =0.09907,~ b = 0.40962,~\tilde b = -0.25347,~\tilde A =0.08318\, ,
\label{eq9}
\end{equation}
and three additional parameters, which are fixed by normalization conditions
\begin{equation}
A = 1.74938,~\bar A = 1.90801,~A_G = 14.27535 \,.
\label{eq10}
\end{equation}

Therefore crucial tests will be provided by measuring flavor and spin 
asymmetries for antiquarks, for which we expect \cite{bour02,bha01} 
\begin{equation}
\Delta \bar u(x) > 0 > \Delta \bar d(x) \, ,
\label{eq12}
\end{equation}
\begin{equation}
\Delta \bar u(x) - \Delta \bar d(x) \simeq \bar d(x) - \bar u(x) > 0 \, .
\label{eq13}
\end{equation}
The inequality $\bar d(x) - \bar u(x) > 0$ has the right sign to agree with the
defect in the Gottfried sum rule \cite{got67}, but not with the trend shown
at high $x$ by the E886 experiment \cite{nus01}. An important test will
be provided by studying $W^{\pm}$ production at RHIC-BNL at
$\sqrt{s} = 200 \mbox{GeV}$ \cite{bour02}.

\begin{figure}
\begin{center}
\leavevmode {\epsfysize= 14.cm \epsffile{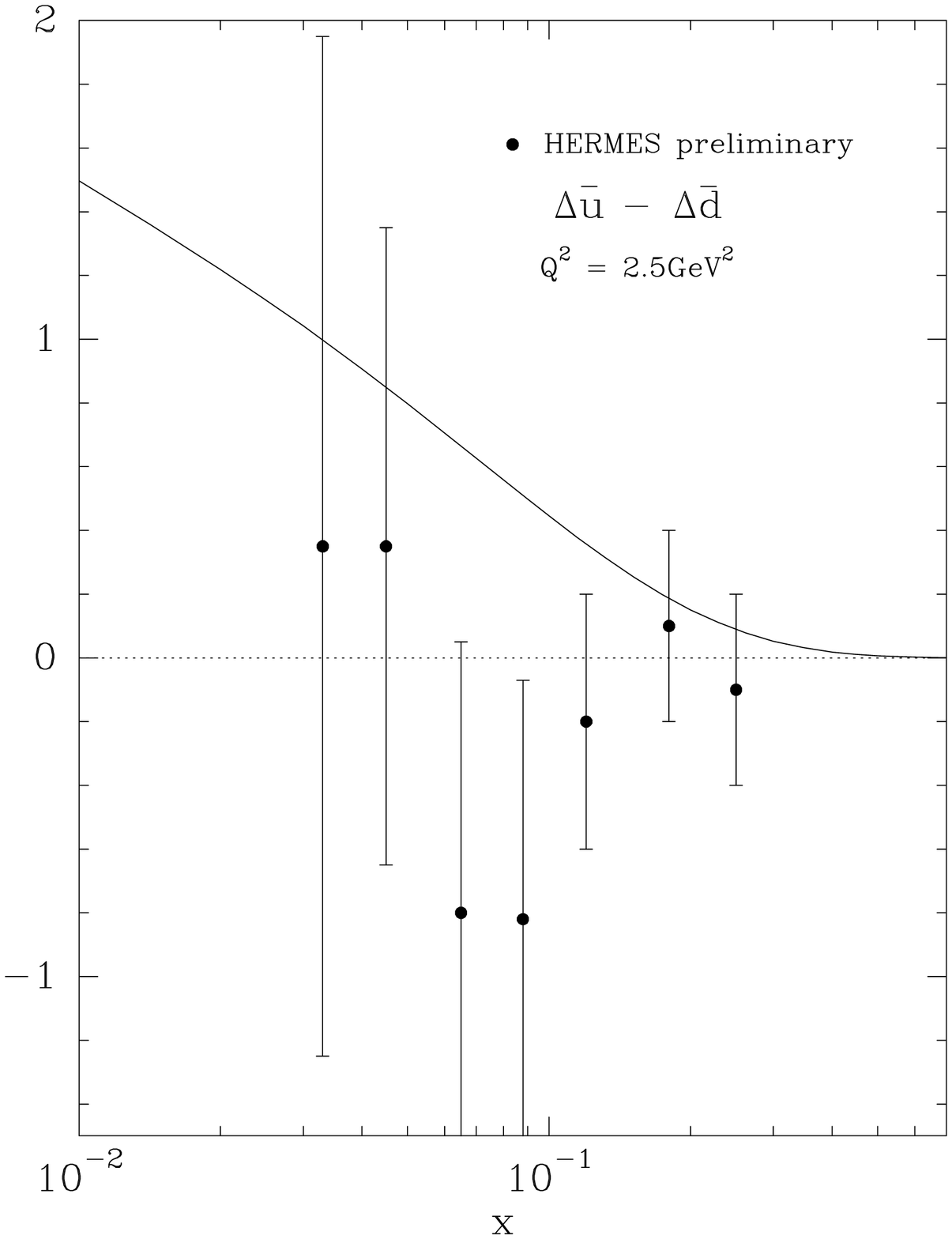}}
\end{center}
\caption[*]{\baselineskip 1pt
Flavor asymmetry $\Delta \bar u -\Delta \bar d$ of the light sea quark
as a function of $x$, for $Q^2 = 2.5\mbox{GeV}^2$. Preliminary data from
HERMES Coll. \cite{herm02}.
}\label{fig1}
\end{figure}

The HERMES Collaboration has provided \cite{herm02} a measurement of
$\Delta \bar u(x) - \Delta \bar d(x)$ displayed in Fig. \ref{fig1},
which shows, within the large errors, consistency both with the vanishing
value implied by a flavor symmetric polarization of the sea and with
our predicted positive value, but disfavors the large positive
values predicted by the chiral QSM \cite{dres00}.

\begin{figure}
\begin{center}
\leavevmode {\epsfysize= 14.cm \epsffile{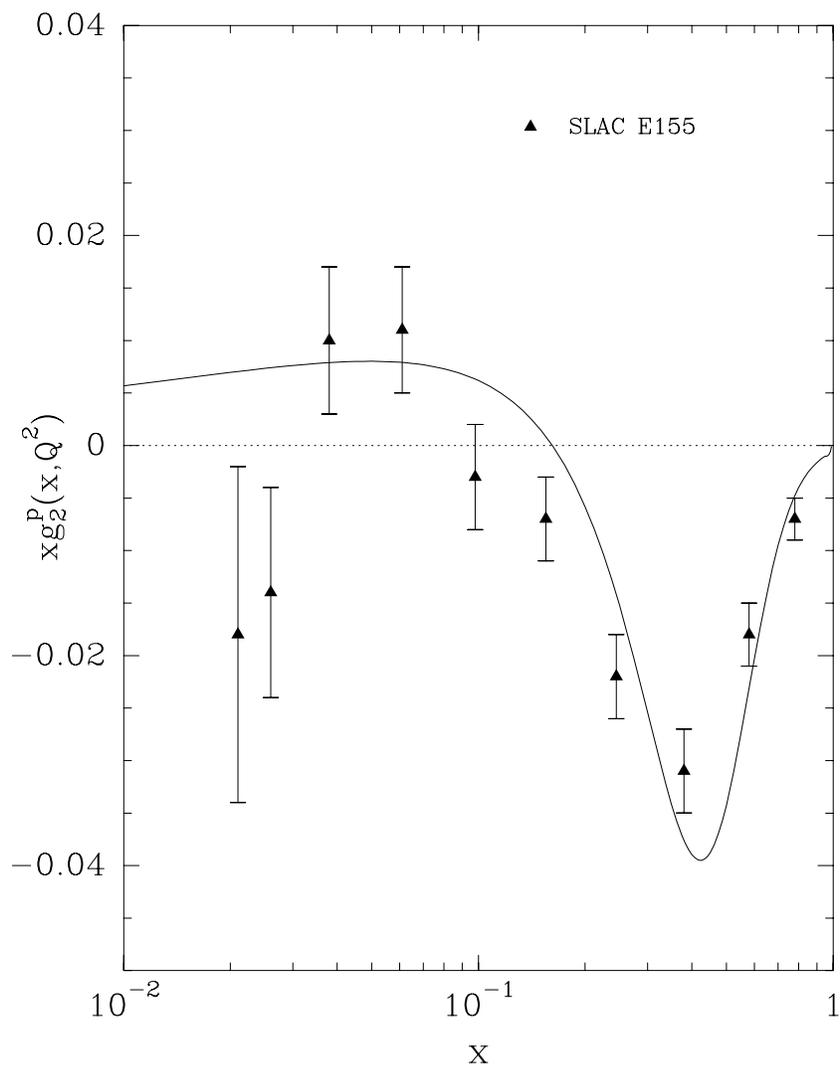}}
\end{center}
\caption[*]{\baselineskip 1pt
$xg_2$ for proton as a function of $x$, for $Q^2=4\mbox{GeV}^2$. Data from
SLAC E155 \cite{slac02}.
}\label{fig2}
\end{figure}

\begin{figure}
\begin{center}
\leavevmode {\epsfysize= 14.cm \epsffile{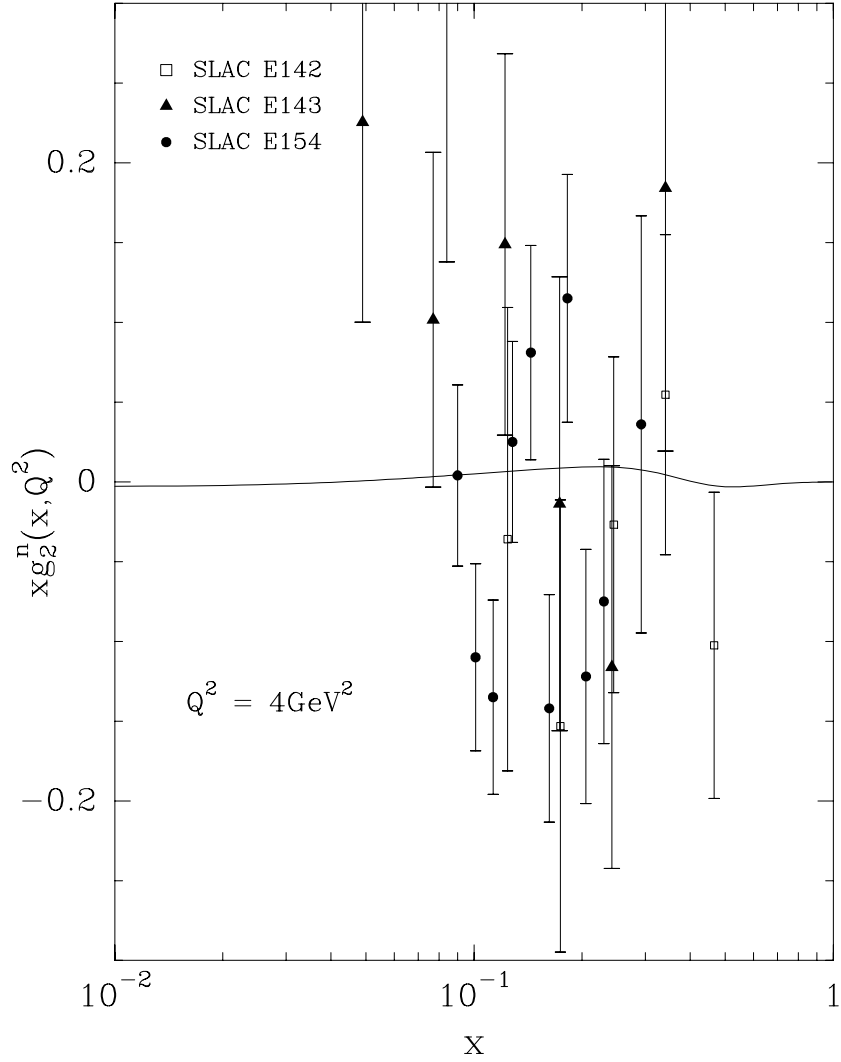}}
\end{center}
\caption[*]{\baselineskip 1pt
$xg_2$ for neutron as a function of $x$, for $Q^2=4\mbox{GeV}^2$. Data from
E142, E143, E154 \cite{e14296}-\cite{e15497}.
}\label{fig3}
\end{figure}

\begin{figure}
\begin{center}
\leavevmode {\epsfysize= 14.cm \epsffile{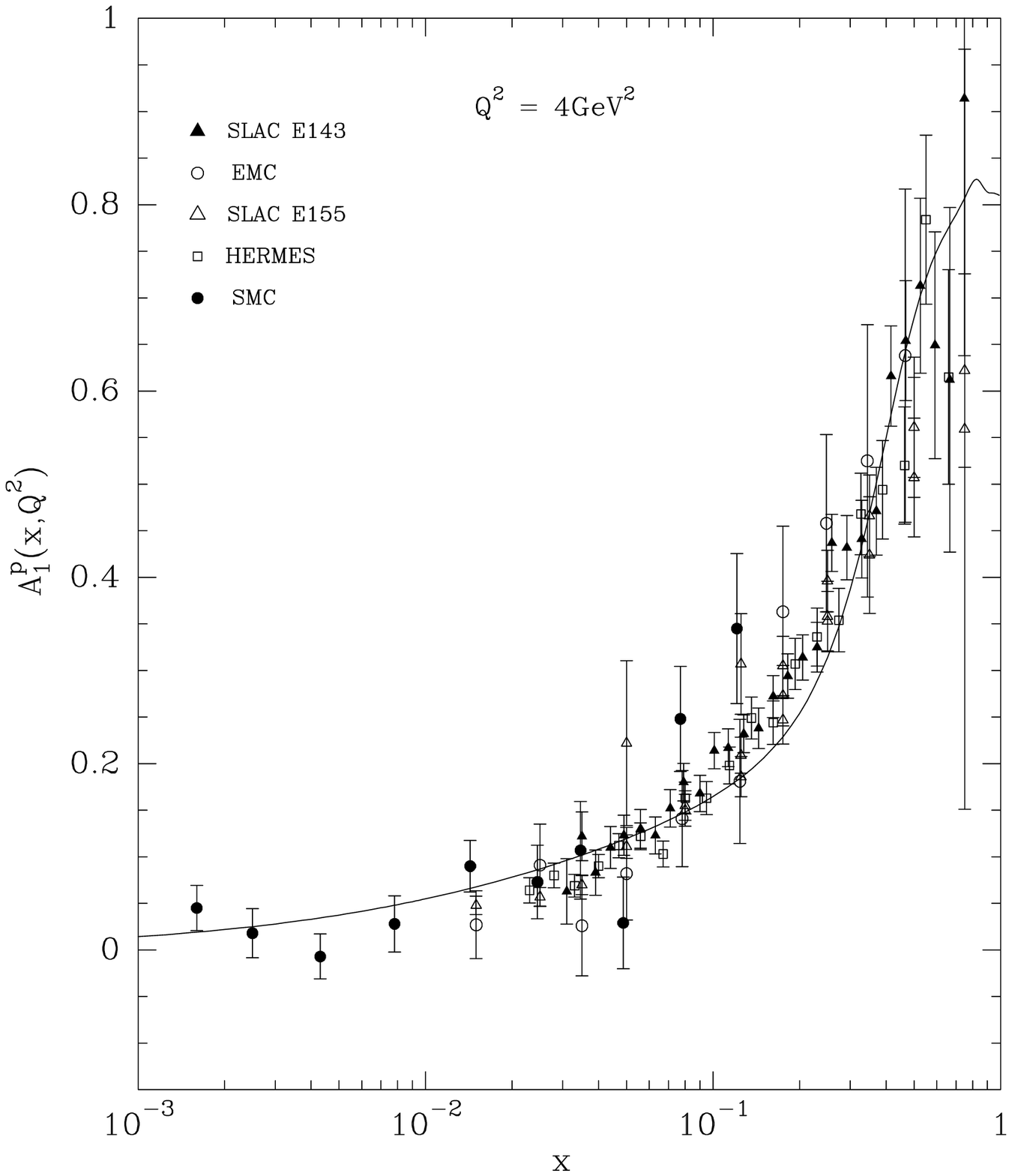}}
\end{center}
\caption[*]{\baselineskip 1pt
$A_1^p$ as a function of $x$, for $Q^2=4\mbox{GeV}^2$.
Data from E143\cite{e14398}, EMC\cite{emc89},
E155\cite{e15500},  HERMES\cite{herm98}, SMC\cite{smc99}.
}\label{fig4}
\end{figure}

\begin{figure}
\begin{center}
\leavevmode {\epsfysize= 14.cm \epsffile{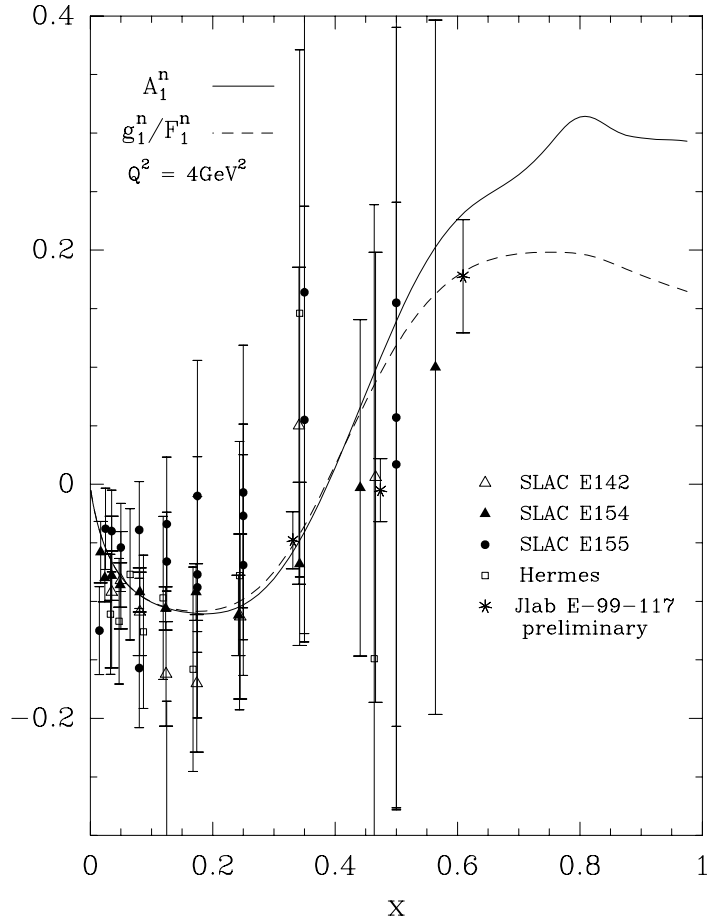}}
\end{center}
\caption[*]{\baselineskip 1pt
$A_1^n$ as a function of $x$, for $Q^2=4\mbox{GeV}^2$ solid curve, $g_1^n/F_1^n$
dashed curve. Data from E142\cite{e14296}, E155\cite{e15500},
E154\cite{e15497b}, HERMES\cite{herm97}, Jlab E-99-117\cite{jlab02}.
}\label{fig5}
\end{figure}

The other polarized structure function $g_2(x,Q^2)$, if one neglects 
twist-three contributions, is given in terms of $g_1(x)$ by the
Wandzura-Wilczek formula \cite{wan00} 
\begin{equation}
g_2^{WW}(x,Q^2) = -g_1(x,Q^2) + \int_x^1\, \frac{g_1(y,Q^2)}{y}dy \, .
\label{eq14}
\end{equation}
We compare our prediction for $g_2^{p,WW}(x,Q^2)$ with the preliminary data
from SLAC \cite{slac02} in Fig. \ref{fig2} and we conclude that the
theoretical curve is in good agreement with the experimental data.
In Fig. \ref{fig3} we compare our prediction for $g_2^{n,WW}(x,Q^2)$ with
the available measurements \cite{e14296}-\cite{e15497} 
and, once again, there is no
disagreement with the data, since about the same number of the central values
of the experimental points, which are affected by large experimental errors, 
fall above and  beneath the theoretical curve.
The functions $g_1(x,Q^2)$ and $g_2(x,Q^2)$ enter in the expression of
the asymmetry $A_1(x,Q^2)$ measured in polarized DIS, as follows
\begin{equation}
A_1(x,Q^2) = {[g_1(x,Q^2) - \gamma^2 g_2(x,Q^2)]\,2 x[1 + R(x,Q^2]
\over (1+\gamma^2)F_2(x,Q^2)} \, ,
\label{eq15}
\end{equation}
where $\gamma^2 = 2M_p x/Q^2$ and $R$ is the ratio of longitudinal and transverse
virtual photo-absorption cross sections. 
 We plot in Figs. \ref{fig4}, \ref{fig5},
our predictions for $A_1^p$ and $A_1^n$ respectively, and compare them with the
available experimental results \cite{e14296}-\cite{herm97},
including three recent points for
$A_1^n$ at high $x$ measured at Jefferson Lab. \cite{jlab02}, which
are in fair agreement with our predictions.
In particular the positive values found at high $x$ for $A_1^n$ agree with 
the dominance of the parton $d^{+} (u^{+})$ at higher $x$ in the 
neutron
(proton), which is a typical consequence of our parton statistical approach.
In Fig. \ref{fig5}, the dashed curve corresponds to $A_1^n=g_1^n/F_1^n$, which is obtained
by making the approximation $g_2^n=-g_1^n$. We see that the use of the exact
Wandzura-Wilczek expression for $g_2^n$ Eq. (\ref{eq14}) leads to a larger $A_1^n$ at
high $x$, as shown by the solid curve. The same effect exists also for the proton case.

\begin{figure}
\begin{center}
\leavevmode {\epsfxsize= 12.cm \epsffile{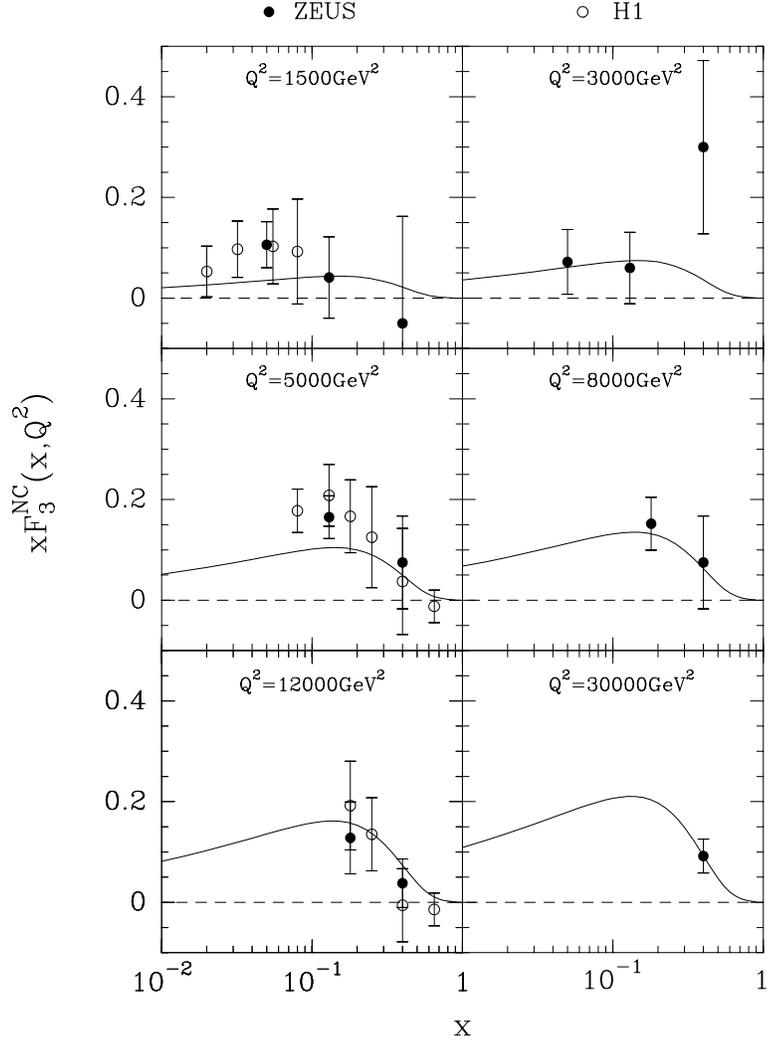}}
\end{center}
\caption[*]{\baselineskip 1pt
The structure function $xF_3^{NC}$ as a function of $x$, for different $Q^2$.
Data from ZEUS Coll. \cite{zeus02}, H1 Coll. \cite{h101}.
}\label{fig6}
\end{figure}

Finally we show in Fig. \ref{fig6} our predictions in very good agreement
with the behavior at high $Q^2$ and large $x$ of
the neutral current structure function $xF_3^{NC}(x,Q^2)$,
measured by ZEUS \cite{zeus02} and H1 \cite{h101}, at HERA in $e^{\pm}p \rightarrow e^{\pm}X$.

To conclude, the above comparison between recent experimental results
and the predictions of the statistical parton distributions, shows that this simple and physical
approach remains reliable. We look forward to more severe tests provided by the flavor
and spin asymmetries of $\bar q$ distributions.

\vskip 0.3cm \noindent {\bf Acknowledgments:} 
We thank J.P. Chen, Z.E. Meziani, C. Vall\`ee and X. Zheng for usefull discussions on recent
data of their collaborations. 

\newpage
\end{document}